\documentclass[sigconf]{acmart}

\newcommand{\quotes}[1]{``#1''}
\usepackage{multirow}

\AtBeginDocument{%
  }

\setcopyright{acmlicensed}
\copyrightyear{2024}
\acmYear{2024}
\acmDOI{XXXXXXX.XXXXXXX}

\acmConference[Conference acronym 'XX]{
  conference title}{June 03--05,
  2018}{Woodstock, NY}
\acmISBN{978-1-4503-XXXX-X/18/06}

\begin{document}

\title{Unlocking the `Why' of Buying: Introducing a New Dataset and Benchmark for Purchase Reason and Post-Purchase Experience }



   \author{Tao Chen}
\email{taochen@google.com}
\affiliation{%
  \institution{Google DeepMind}
  \city{Mountain View}
  \state{CA}
  \country{USA}
}
 
 \author{Siqi Zuo}
\email{siqiz@google.com}
\affiliation{%
  \institution{Google}
  \city{Mountain View}
  \state{CA}
  \country{USA}
}

   \author{Cheng Li}
\email{chgli@google.com}
\affiliation{%
  \institution{Google DeepMind}
  \city{Mountain View}
  \state{CA}
  \country{USA}
}
 
    \author{Mingyang Zhang}
\email{mingyang@google.com}
\affiliation{%
  \institution{Google DeepMind}
  \city{Mountain View}
  \state{CA}
  \country{USA}
}
 
  \author{Qiaozhu Mei}
  \authornote{The work was done as a visiting researcher at Google DeepMind.}
\email{qmei@umich.edu}
\affiliation{%
  \institution{University of Michigan}
  \city{Ann Arbor}
  \state{MI}
  \country{USA}
}

     \author{Michael Bendersky}
\email{bemike@google.com}
\affiliation{%
  \institution{Google DeepMind}
  \city{Mountain View}
  \state{CA}
  \country{USA}
}
 




\begin{abstract}


In business and marketing, analyzing the reasons behind buying is a fundamental step towards understanding consumer behaviors, shaping business strategies, and predicting market outcomes.
Prior research on purchase reason has relied on surveys to gather data from users. However, this method is limited in scalability, often focusing on specific products or brands, and may not accurately represent the broader population due to the restricted number of participants involved.

In our work, we propose  \emph{purchase reason prediction} as a novel task for modern AI models. To benchmark potential AI solutions for this new task, we first generate a dataset that consists of real-world explanations of why users make certain purchase decisions for various products.  Our approach induces LLMs to explicitly distinguish between the reasons behind purchasing a product and the experience after the purchase in a user review. An automated, LLM-driven evaluation as well as a small scale human evaluation confirm the effectiveness of this approach to obtaining high-quality, personalized purchase reasons and post-purchase experiences. With this novel dataset, we are able to benchmark the purchase reason prediction task using various LLMs. Moreover, we demonstrate how purchase reasons can be valuable for downstream applications, such as marketing-focused user behavior analysis,  post-purchase experience and rating prediction in recommender systems, and serving as a new approach to justify recommendations. 

\end{abstract}



\maketitle


\section{Introduction}
\label{sec:introduction}




\begin{figure}[th]
    \centering
    \includegraphics[width=5.5cm]{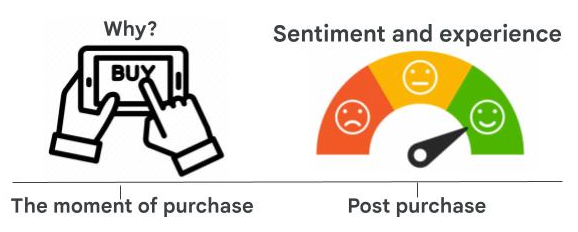}
    \caption{The temporal relationship between purchase reason and post-purchase experience.}
    \label{fig:temporal}
\end{figure}

In the fields of business and marketing, analyzing the reasons behind purchasing is a fundamental step towards understanding consumer behaviors, shaping business strategies, and predicting market outcomes~\cite{chang1994price, cobb1995brand,bleize2019factors,Svatosova2013}.  These reasons include personal needs, preferences, brand perception, trust, perceived value, etc. Traditionally, business and marketing research has relied on questionnaires or interviews to gather information about customer purchase reasons for specific products or brands. This approach typically asks participants to select from a pre-defined list of reasons or to write in their own. While informative, these methods have limitations in terms of scalability and may not accurately reflect the wider population due to the limited number of participants.

To address these limitations, we propose a novel \textit{purchase reason prediction} task, for which one may charge an AI model to generate why a user could purchase an item. To facilitate this task, we build a real-world purchase reason dataset from user reviews that cover various products.  However, user reviews, written after a purchase, can reveal both the initial reasons for buying (i.e., purchase reason) and how the user feels about the product after using it (i.e., post-purchase experience).  Figure~\ref{fig:temporal} illustrates the temporal relationship between the two concepts. While some existing research also uses reviews to mine explanations for recommender systems, they focus on how the item was rated (post-purchase) instead of why it was bought in the first place. One study~\cite{Li2021} extracts the most commonly occurred near-duplicate sentences across reviews, resulting in short and generic comments about an item (e.g., \quotes{\textit{Excellent movie}}). Others extract review sentences~\cite{Geng2022} or segments~\cite{Ni2019} that mention one or more pre-selected item features/aspects (e.g., ``\textit{The quality of the material is great}''). Since reviews are written after a purchase, user’s sentiments towards the item are primarily based on post-purchase user experience. Table~\ref{tab:example_review} is an example contrasting our proposed tasks and the existing post-purchase experience tasks.

\begin{table}[t]
    \centering
    \caption{An example compares existing tasks with our proposed purchase reason and  post-purchase experience task.}
    \label{tab:example_review}
    \small
    \begin{tabular}{p{0.95\linewidth}} \hline
         \underline{Product}: \\ Google Pixel 8 - Unlocked Android Smartphone \\
         \underline{Review}: \\ I bought this phone as a birthday gift for my teenage daughter who is a fan of AI features. My daughter loves the AI photo editor, with which she successfully removed a stranger from our recent family reunion photo. I highly recommend this phone. \\ \hline
         (\textit{\textbf{proposed new task}}) Purchase reason: Birthday gift for a teenage daughter who likes AI features. \\
         (\textit{\textbf{proposed improved task}}) Post-purchase experience: The daughter loves the AI photo editor and found it a useful tool. Highly recommend. \\
          (\textit{existing task}) Common snippet based experience: highly recommend. \\
         (\textit{existing task}) Feature based experience: AI photo editor. \\
          \hline
          
    \end{tabular}
\end{table}

In our work, we propose to explicitly distinguish between the reasons behind purchasing a product and the experience after the purchase in a user review. Our approach leverages a large language model (LLM) to generate a dataset from user reviews, which jointly captures purchase reasons and a highly relatable, personalized post-purchase experience. We propose three dimensions to measure the data quality and validate the effectiveness of our approach through automated LLM evaluator and small-scale human feedback. The resulting dataset is a high-quality, personalized set of purchase reasons and post-purchase experience. With this dataset, one can develop models to generate user predictions from two aspects,  the relevance to the user need (i.e., purchase reason) and the preference for the particular item (i.e., post-purchase experience).  We benchmark the dataset against the two tasks with various LLMs and explore the effectiveness of different user and item representations.

Our proposed purchase reason prediction task and the curated dataset have multiple applications. We demonstrate its practical value in marketing-focused user behavior analysis. Moreover, we argue that understanding the purchase reasons is crucial for personalized recommender systems, especially because recommendations are made \textit{before} the user purchases an item. They reveal the user's personal information needs and motives, often beyond the particular reviewed item. We validate its effectiveness in improving post-purchase experience generation and rating prediction. Further, we suggest purchase reasons could be a new way to explain recommendations, showing how they connect to user needs, and complement existing explanations based on post-purchase experience (preference for the item itself). 

\begin{figure}[th]
    \centering
    \includegraphics[width=5.5cm]{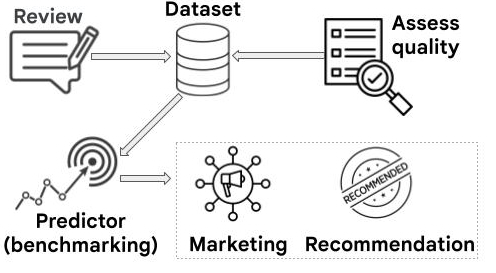}
    \caption{The overview of our work. We construct the dataset from user reviews,  validate its quality (Sec.~\ref{sec:dataset_construction} and \ref{sec:data_analysis}),  build prediction models (Sec.~\ref{sec:experiment}), and demonstrate its usefulness in various applications (Sec.~\ref{sec:applications}). }
    \label{fig:pipeline}
\end{figure}

Figure~\ref{fig:pipeline} provides the overview of our work. 
To summarize,  the main contributions of this work are as follows:
\begin{itemize}
    \item To the best of our knowledge, we are the first to propose the task of purchase reason prediction.  
    \item We propose a simple yet effective LLM-based approach to generate a high-quality, personalized dataset consisting of both purchase reasons and post-purchase experiences (Sec.~\ref{sec:dataset_construction}).
     \item  We benchmark the tasks of purchase reason and post-purchase experience generation in the context of e-commerce (Sec.~\ref{sec:experiment}).
    \item We demonstrate the usefulness of purchase reasons in various downstream tasks, including user behavior analysis, post-purchase experience and ratings prediction and serving as a new explanation in recommender system (Sec.~\ref{sec:applications}).
\end{itemize}
\section{Related Work}
\label{sec:related_work}

We first review prior work on purchase reason, followed by dataset construction and methods for personalized explanations in recommender systems. We then discuss the use of LLMs for text generation and evaluation.

\subsection{Purchase Reason}
\label{subsec:related_work_purchase_reason}

Customer purchase intention reflects how likely a potential customer is to buy a product or a service~\cite{chang1994price}.  This intention is shaped by a combination of rational and emotional factors, such as personal needs, preferences, brand perception, trust, perceived value,  often referred to as the \quotes{purchase reason}, which ultimately guide the buying decision~\cite{cobb1995brand,bleize2019factors,Svatosova2013}. Traditionally, business and marketing research has used surveys or interviews to gather information about customer purchase reasons for specific products or brands (e.g., \cite{Hamari2017,Ozaki2011,Rana2020}). This approach typically involves providing a pre-defined list of reasons or asking participants to write their own. 
While informative, these methods have limitations in terms of scalability and may not accurately reflect the broader population due to the limited number of participants.

Our research addresses these limitations by leveraging LLMs to automatically extract user's self-reported purchase reasons from user reviews, thereby creating a personalized purchase reason dataset.  These purchase reasons are provided directly by users themselves, and representative of a large user pool.  Utilizing this dataset, we are the first to develop a machine learning model capable of predicting purchase reasons for buyers or potential buyers in an e-commerce environment.

\subsection{Personalized Explanation Dataset}
\label{subsec:related_work_exp_dataset}
A related line of research is creating personalized explanation dataset for recommender system. One work~\cite{Colas2023} uses product description as ground truth explanation, which is fact-grounded but not personalized, while the majority  mine personalized explanations from user reviews or Tips (a concise form of reviews only on Yelp mobile)~\cite{Li2017}. These work extracts various information (e.g., aspect-specific sentences) from user reviews~\cite{Chen2018, Li2017, Chen2019, Wang2018, Li2020, Geng2022, Ni2019}.
They discard segments that are too personal (containing first-person or third-person pronouns) or too short/long. \citet{Li2021} extract commonly occurred near-duplicate sentences as explanations, which are often short and generic. 

The above approaches capture customer satisfaction rather than purchase motives because they do not distinguish pre- from post-purchase experiences. To our knowledge, we are the first to extract purchase reasons directly from user reviews.

\subsection{Personalized Explanation Generation}
\label{subsec:related_work_exp_generation}
Existing approaches, such as ranking-based~\cite{Li2021, Li2023b} and template-based~\cite{Zhang2024b, Wang2018b}, tend to produce generic explanations with limited language flexibility and personalization. Natural language generation (NLG) is widely used to generate free-text explanations. Early studies fine-tuned seq2seq models like LSTM~\cite{Costa2018}, GRU~\cite{Li2017, Ni2019}, Transformer~\cite{Li2021b}, T5~\cite{liu2023} or GPT-2~\cite{Li2023}. The results were often generic and not fluent~\cite{liu2023}. Recent work with powerful LLMs (e.g., ChatGPT~\cite{liu2023}) in zero or few-shot setups has significantly improved quality.
However, effectively representing users and items remains a challenge for NLG-based methods, with most studies relying on textual descriptions or IDs~\cite{liu2023,Li2023}, with the latter limiting generalization to new IDs.

A shared challenge of all these approaches is systematically evaluating generated personalized explanations.  In our work, we introduce purchase reason as a potential new form of explanation, and we collect a corresponding dataset and benchmark to facilitate research in this area.

\subsection{LLMs for Text Generation and Evaluation}
\label{subsec:related_work_llm}

While LLMs demonstrate exceptional capabilities in generating text~\cite{Zhao2023}, their tend to hallucinate~\cite{Zhang2023, Huang2023}. LLMs have been widely used to evaluate the quality of generated text across various domains~\cite{Zhao2023, Wang2023b, Kocmi2023, Wang2023, Zhang2024}, often achieving state-of-the-art or competitive correlation with human judgments. Our work also leverages LLMs as a core tool, employing them to extract purchase reasons from user reviews for dataset creation, evaluate the quality of the dataset, and build a model for predicting purchase reasons.
\section{Creating a Purchase Reason Dataset}
\label{sec:dataset_construction}


As we discussed in Section~\ref{subsec:related_work_purchase_reason}, existing business and marketing research often relies on surveys to understand customer motivations.  However, these works focus on specific products or brands, and their findings may not be generalizable due to small sample sizes. Our work takes a different approach, directly mining customer reviews to identify self-reported purchase reasons. We recognize that reviews can include both the initial motivations for buying and the user's feelings after using the product, so we carefully differentiate between these two types of information. The key distinction lies in the timing: purchase reasons precede the purchase and explain the decision to buy, while post-purchase experiences occur after the purchase and influence the user's rating. While other research has used reviews to extract personalized explanations in recommendation (Section~\ref{subsec:related_work_exp_dataset}), these efforts primarily focus on extracting user sentiment towards the item, which is typically a reflection of post-purchase experience.

As LLMs have shown strong performance in many NLP tasks, we propose to use LLMs to extract both purchase reasons and post-purchase experiences from user reviews, rather than adopting the traditional extractive methods used by prior recommendation explanation works. We carefully devise strategies to improve generation quality and combat hallucinations. Moreover, we use LLMs as auto-raters to assess the generations from multiple aspects and further improve data quality.  
Specifically, we demonstrate the effectiveness of our solution on the Amazon product review dataset
\cite{Ni2019} employing a diverse range of LLMs. These include two commercial models, Gemini Ultra~\cite{team2023gemini} and GPT-4 Turbo~\cite{openai2023gpt4}, as well as an open-source LLM Gemma-7B~\cite{gemmateam2024}. We detail our end-to-end solution below.


\subsection{Extracting Purchase Reason using LLMs}

Given a product and an associated user review, our goal is to use an LLM to 
\textit{extract} this user's purchase reason and post-purchase experience. While we conceptualize this task as extraction, the LLM's actual function is to  identify and summarize relevant information.

Initially, we treated the two generation tasks separately but encountered two issues. First, LLMs struggle to distinguish between the two tasks, leading to overlapping or miscategorized generations. Second, when reviews lack explicit purchase reasons, the model either infers them from product information or hallucinates. We address these issues with strategies detailed below, with our final prompt in Table~\ref{tab:data_prompt}.

1. We instruct the model to perform the two generation tasks simultaneously in one prompt, enforcing a clear distinction between purchase reasons and post-purchase experiences.

2. To reduce hallucination, we separate purchase reasons into \textit{explicit} (directly stated in the review) and \textit{implicit} (inferred from the review and product information). The model can leave either blank if information is not found.  These are combined in later evaluation and analysis.

3.  We require the model to provide supporting evidence for its generated purchase reasons. This further helps combat hallucination and promotes more accurate and concise generations.

4. We incorporate two in-context examples: one demonstrating both purchase reason and post-purchase experience, and the other featuring a concise review containing only post-purchase experience. We find that these examples effectively mitigate the generation of false purchase reasons and enhance the completeness of post-purchase experiences, especially for short reviews.

\begin{table}[t]
    \caption{Prompt snippet used to extract purchase reason and post-purchase experience from user reviews.}
    \label{tab:data_prompt}
    \small
    \centering
    \begin{tabular}{|p{0.95\linewidth}|}
    \hline
You are a customer engagement specialist at Amazon. Given a product and a customer review, please identify the customer's purchase reason and post-purchase experience:\\
1. explicit\_purchase\_reason:The reason for purchasing the product that is directly mentioned in the review. Describe in detail the thought processes before the purchase. Leave null if not mentioned.\\
2. implicit\_purchase\_reason: The reason for purchasing the product that is not explicitly mentioned in the review but can be inferred from the review and the product description. Leave null if not mentioned.\\
3. purchase\_reason\_explanation: Briefly justify your reasoning behind the identified explicit and implicit purchase reasons.\\
4. post\_purchase\_experience: Summarize in 2-3 lines how the product met (or didn't meet) the customer's expectations, based on the review. \\\\

Please be as specific and relating to the customer's personal context as much as possible.\\ 
... \\
\hline
    \end{tabular}
\end{table}

\subsection{Rating Purchase Reason using LLMs}
\label{subsec:auto_rater}
Prior work has demonstrated a great success of automatic evaluation of text generation with LLMs (detailed in Section~\ref{subsec:related_work_llm}). Inspired by this, we explore the usage of LLMs in assessing the quality of our dataset from both precision and recall perspectives. More specifically, we present the two generations (purchase reason and post-purchase experience) in our dataset alongside the original review from which they were derived. We then ask LLMs to judge the quality of the generations based on the following three dimensions:

\textbf{Hallucination}: if the generation contains any completely irrelevant information that are not described or implied in the product information or the user review. This metric is designed to measure the \textbf{precision} of the generation.

\textbf{Type Match}:
if the model accurately distinguishes between the two types of generated text: purchase reasons and post-purchase experiences. We observe that the model sometimes confuse post-purchase experience as purchase reason, but rarely the opposite.  This metric also focuses on \textbf{precision}, ensuring the model correctly categorizes the generated text.

\textbf{Completeness}: if the generation covers all relevant aspects present in the  review and the product description. This metric is focused on \textbf{recall}, ensuring the generated text does not miss any relevant details.

To simplify the task, we conduct two separate evaluations for purchase reason and post-purchase experience.  We further leverage the three evaluation results to filter out noisy generations, aiming at improving the dataset quality.

\section{Dataset Evaluation and Analysis}
\label{sec:data_analysis}


We use the Amazon product review 5-core dataset for experiments. The full 5-core dataset consists of reviews from all users and items that have at least 5 reviews, resulting in 75 million reviews in total. Considering the cost of using LLMs, we randomly sample 10K reviews for experiments. 
We apply the LLM extractor to construct the purchase reason and post-purchase experience dataset and utilize the LLM auto-rater to judge the data quality. 
In the following, we first validate the effectiveness of our LLM auto-rater and then describe the quality and characteristics of the generated  dataset. We choose Gemini Ultra as the primary LLM for both extractor and auto-rater, but also  assess the generalizability of our approach using GPT-4 Turbo and the more compact, open-source Gemma-7B.

\subsection{Effectiveness of LLM Auto-rater}
\label{subsec:human_eval}
To evaluate our Gemini Ultra auto-rater, we conduct a small scale human annotations following the same guideline used by the auto-rater. In the pilot study (20 reviews, 4 annotators), agreement was high: perfect consensus on hallucination, and 90\% agreement on completeness and type match. Disagreements were resolved through discussion. Subsequently, a larger set of 100 randomly sampled reviews are annotated (one annotator per review).

Table~\ref{tab:auto_rater_human} shows that the LLM auto-rater is highly effective, achieving almost perfect correlation with human judgment for hallucination and type match. Agreement on completeness reached 92\% for purchase reasons and 95\% for post-purchase experiences. Analysis suggests the auto-rater is stricter than humans for post-purchase experience completeness but more lenient for purchase reason completeness, indicating that identifying purchase reasons may be a more challenging task.

\begin{table}[h]
    \caption{The agreement rate between the LLM auto-rater and human annotator on 100 randomly sampled reviews.}
    \small
    \label{tab:auto_rater_human}
\begin{tabular}{l|ccc} \hline
           & Hallucination & Completeness & Type Match        \\ \hline
Purchase Reason     & 99\%         & 92\%         & \multirow{2}{*}{100\%} \\
Post-Purchase Experience & 100\%         & 95\%         &                        \\ 
\hline
\end{tabular}
\end{table}

\subsection{Quality of the Dataset}
\label{subsec:eval_result}

Table~\ref{tab:rater_result} shows the evaluation of the generated 10K dataset using the LLM auto-rater. Top product categories are also shown for understanding quality variations. The LLM generator achieves a low hallucination rate (0.9\% for both purchase reasons and post-purchase experiences) and rarely confuses the two generation tasks (99.1\% are correct). 
The completeness rate is lower than other two dimensions (89.4\% for purchase reasons, 95.9\% for post-purchase experience). As discussed in Sec.~\ref{subsec:human_eval}, the strict auto-rater considers subtle aspects for the completeness evaluation, suggesting that the actual rate might be higher.

Within the five product categories, ``Sports \& Outdoors'' has the highest completeness rate for both purchase reason (91.3\%) and post-purchase experience (97.1\%). Fashion (``Clothing, Shoes \& Jewelry'') has the lowest rate (86.9\%) for purchase reasons, while ``Books'' (95.5\%) has the lowest for post-purchase experiences. This discrepancy might be due to varying purchase and review behaviors across categories. Unlike ``Sports \& Outdoors'' reviews, where buyers often state their motivations explicitly, reviews for clothing, shoes, or jewelry may require the LLM to infer motivations from subtle cues, potentially leading to omissions. Conversely, the descriptive language in book reviews, compared to the brevity of ``Sports \& Outdoors'' reviews, poses a challenge for the LLM to capture the nuances of reader sentiment.

\begin{table}[h]
     \caption{The LLM auto-rater results on 10K dataset. Elec denotes \quotes{Electronic}.}  
    \label{tab:rater_result}
    \small
\begin{tabular}{l|cccccc} \hline
\multicolumn{1}{l|}{}                    & All      & Fashion & Sports & Books   & Elec. & Home \\ \hline

 \multicolumn{2}{l} {Purchase Reason} \\

Hallucination & 0.9\% & 0.6\% & 1.2\% & 1.2\% & 1.1\% & 1.0\% \\
Completeness    & 89.4\%  & 86.9\% & 91.3\% & 90.0\%& 90.1\%     & 88.1\%            \\\hline
 \multicolumn{2}{l} {Post-purchase Experience} \\
Hallucination  & 0.9\% & 0.5\% & 0.6\% & 0.8\% & 1.7\% & 0.8\% \\
Completeness  & 95.9\% & 96.0\% & 97.1\% & 95.5\% & 96.4\% & 95.8\% \\ \hline
Type Match  & 99.1\% & 99.2\% & 98.9\% & 99.1\% & 98.9\% & 98.8\% \\
\hline
\end{tabular}
\end{table}

\subsection{Dataset Characteristics and Comparisons}

Our LLM extractor successfully identifies purchase reason in 93.94\% of reviews. This includes  36.8\% of reviews with explicitly stated reasons and 84.2\% where reasons are inferred from the review and the product information. The majority of reviews lacking an identified purchase reason are short (55\% are shorter than 20 tokens), suggesting that users may not mention or provide any clues regarding their motivation for buying in these cases. Our LLM identifies post-purchase experience in nearly all reviews (99.93\%). This is because users typically express their satisfaction or dissatisfaction in the reviews, even in very short ones.

\begin{table}[th]
    \centering
    \caption{Examples from our dataset compared to two existing explanation dataset P5~\cite{Geng2022} and EXTRA~\cite{Li2021}. Note that P5 and EXTRA cover different product categories, so we can't provide examples that overlap across all three datasets.}
    \label{tab:example_explanations}
    \small
    \begin{tabular}{p{0.95\linewidth}} \hline

        \textbf{Example 1} \\
         \underline{Product}: Fisher Price Exclusive Medical Kit Pink \\
         \underline{Review}: Such a cute kit. I bought this for my 3 year old and she loves it!  I don't understand the complaints about this product that some were talking about it the reviews. My daughter and I had no problem fitting everything in the case.  We have had this for 5 months now and not a single thing has broken.  My kids love playing with it and love playing veterinarian with the dogs.  I think this is a great product and totally recommend it. \\
         \underline{Purchase reason}:To provide a fun and engaging play experience for the customer's 3-year-old daughter. \\
         \underline{Post-purchase experience}: The medical kit exceeded the customer's expectations. It has remained intact for five months and provides a fun and engaging play experience for the customer's daughter and their pets.\\ 
         \underline{P5 explanation}: I think this is a great product and totally recommend it. \\
         \hline
        
         
         \textbf{Example 2} \\
         \underline{Product}: Bridget Jones's Diary \\
         \underline{Review}: Zellweger plays a perfect single English woman.... looking for love in all the wrong places.  The story reminds me of Pride and Prejudice a little, especially with Colin Firth playing Mr. Darcy.  A very cute, entertaining movie. \\
         \underline{Purchase reason}: Interest in romantic comedies and resemblance to Pride and Prejudice \\
         \underline{Post-purchase experience}: The movie met the customer's expectations, providing a cute and entertaining experience.\\ 
         \underline{EXTRA explanation}: Looking for love in all the wrong places \\
         \hline

    \end{tabular}
\end{table}

Table~\ref{tab:example_explanations} shows purchase reason and post-purchase experience for two reviews in our dataset, along with examples from two previous explanation datasets: EXTRA~\cite{Li2021}, which utilizes commonly occurring near-duplicate phrases as explanations, and P5~\cite{Geng2022}, which extracts sentences containing product features as explanations.

We further characterize our dataset from linguistic aspects (detailed in Table~\ref{tab:dataset_analysis}).
On average,  purchase reasons (11.39 words) are shorter than post-purchase experiences (22.19 words), and both are significantly shorter than the original reviews (63.79 words). As expected, explanations in the EXTRA dataset are very short (4.96 words), while P5 explanations are longer (25.33 words). The type-to-token ratio (i.e., unique word count/total word count)~\cite{Templin1957} shows that our generations have significantly higher vocabulary diversity (0.140 for purchase reason and 0.048 for post-purchase experience) compared to the EXTRA (0.006) and P5 (0.014) datasets.

\begin{table}[th]
    \caption{The linguistic characteristics of our 10k dataset and two other explanation datasets.} 
      \label{tab:dataset_analysis}
    \small
\begin{tabular}{l|ccc} \hline
                         & Avg word count &  Type-to-token\\\hline
Product      & 156.33                        & 0.058                                                            \\
User review              & 63.79                           & 0.053                                                       \\ \hline
Purchase reason & 11.39                          & 0.140                                                           \\
Post-purchase experience & 22.19                          & 0.048                             \\ \hline 
EXTRA~\cite{Li2021}       & 4.96            & 0.006      \\
P5~\cite{Geng2022}        & 25.33            &  0.014  \\ 
\hline                           
\end{tabular}
\end{table}

\subsection{Generalization of LLM Extractor}
To assess the broader applicability of our proposed LLM-based extractor, we evaluate its performance using GPT-4 Turbo ((gpt-4-0125-preview) and Gemma-7B (1.1) on 1k randomly sampled reviews.  The quality of the extracted purchase reason and post-purchase experience data, as measured by the Gemini auto-rater (Table~\ref{tab:data_quality_llms}), reveals that larger models generally perform comparably for all the metrics. The smaller Gemma model is more prone to hallucination, introducing irrelevant information in 8.4\% and 5.8\% of purchase reason and post-purchase experience extractions, respectively. Additionally, Gemma's post-purchase experience completion rate (85.4\%) falls behind the other two models (94.3\% for Gemini and 95.9\% for GPT-4).

\begin{table}[th]
\small
\caption{The auto-rater results on 1000 datasets generated by different models.  Hall./Compl./Type. denotes hallucination/completeness/type match, respectively.}
\label{tab:data_quality_llms}
\begin{tabular}{llllll} \hline
       & \multicolumn{2}{c}{Purchase Reason} & \multicolumn{2}{c}{Post-purchase Experience} &            \\ \hline
Model  & Hall.            & Compl.            & Hall.         & Compl.         & Type.  \\ \hline
Gemini & 1.8\%             & 80.7\%          & 0.7\%                 & 94.3\%               & 98.2\%     \\
GPT-4  & 1.7\%             & 79.1\%          & 0.8\%                 & 95.9\%               & 97.9\%     \\
Gemma  & 8.4\%             & 75.2\%          & 5.8\%                 & 85.4\%               & 90.2\%    \\\hline
\end{tabular}
\end{table}
\section{A Benchmark of Purchase Reason Prediction}
\label{sec:experiment}


We benchmark our newly developed dataset in the novel generation task -- predicting purchase reason and post-purchase experience,  within the e-commercial setting.  More formally, we define the task as the following: Given information about a user and an item, generate why the user could purchase the item (i.e., purchase reason) and what's the experience after using the item (i.e., post-purchase experience). The item could be one the user has already bought or one that has been recommended to them and they are considering purchasing. This task setup is similar to the existing explanation generation in recommender systems~\cite{liu2023, Geng2022}.

We would like to answer three research questions:
\begin{itemize}
    \item \textbf{RQ 1}: Which user and item signals are most effective?
    \item \textbf{RQ 2}: Can mainstream LLMs perform well with zero/few-shot prompting?
    \item \textbf{RQ 3}: How can we enhance the performance of purchase reason predictions?
\end{itemize}

In the following, we first describe our experimental setup, followed by conducting comprehensive experiments to answer the three research questions.
We then analyze our findings, discuss the challenging of this task, and suggest potential directions for model improvement, with the hope this dataset will serve as a valuable resource to inspire further research in this area.

\subsection{Experiment setup}

\textbf{Dataset.} We use our newly developed dataset for experiments. The original dataset only includes review-product pairs, so we enrich it by adding historical review data related to both the user and the item. More specifically, we include reviews written by the user before purchasing the target item, as well as reviews written by other users prior to the target user's purchase. 

\textbf{Models.} We again choose LLMs with zero/few-shot setup 
as the tasked models for benchmarking, given LLMs have demonstrated superior performance in text generation across various tasks (Sec.~\ref{subsec:related_work_llm}). For instance, LLM-generated explanations are significantly favored by human raters over small pre-LLM models~\cite{liu2023}.
Our primary experimental LLM is Gemini Ultra, employed in a zero-shot setup. Additionally, we benchmark against GPT-4 Turbo, open-source Gemma-7B, and a pre-LLM state-of-the-art recommendation explanation generation model P5 ~\cite{Geng2022} for comparisons. 

\textbf{Evaluation metrics.} We evaluate model generated purchase reason and post-purchase experience against the ``ground-truth'' provided in our dataset using commonly adopted text generation evaluation metrics. These include BLEU-1, BLEU-2~\cite{papineni2002bleu}, ROUGE-1, ROUGE-2, and ROUGE-Lsum~\cite{lin2004rouge} to assess lexical similarity, and BERTScore~\cite{Zhang2020} to measure  semantic similarity. 

\subsection{User and Item Representation (RQ 1)}
\label{subsec:user_item_representation}

\begin{table*}[h]
\small
\caption{Performance of Gemini Ultra using different user and item representations (\%). B/R/RL/BERT denotes BLEU, ROUGE, ROUGE-Lsum, BERTScore respectively.}
\label{tab:user_item_representaiton}
\begin{tabular}{l|rrrrrr|rrrrrr} \hline
                                             & \multicolumn{6}{c}{Purchase Reason}                                                                                                                                          & \multicolumn{6}{c}{Post-purchase Experience}                                                                                           \\
Method                                       & \multicolumn{1}{c}{B1}      & \multicolumn{1}{c}{B2}     & \multicolumn{1}{c}{R1}      & \multicolumn{1}{c}{R2}     & \multicolumn{1}{c}{RL}   & \multicolumn{1}{c}{BERT} & B1   & \multicolumn{1}{c}{B2} & \multicolumn{1}{c}{R1} & \multicolumn{1}{c}{R2} & \multicolumn{1}{c}{RL} & \multicolumn{1}{c}{BERT} \\ \hline
{UserReview-ItemReview} & 17.7                        & 8.8                        & 22.8                        & 8.9                        & 21.1                        & 25.5                     & 21.8 & 8.7                    & 24.2                   & 5.4                    & 20.0                      & 31.3                    \\
\textbf{UserReview-Item}                     & { 18.1} & { 9.4} & {23.4} & { 9.8} & { 21.7} & 26.1                     & 23.0 & 8.7                    & 24.3                   & 5.4                    & 20.2                      & 29.3                     \\
ItemReview                                   & 13.0                        & 6.5                        & 20.8                        & 7.7                        & 19.4                        & 23.2                     & 22.2 & 9.2                    & 23.9                   & 5.9                    & 19.7                      & 30.5                     \\
UserReviewSummary-ItemReviewSummary                      & 5.6                         & 0.8                        & 6.5                         & 0.7                        & 6.2                         & 7.3                      & 19.6 & 5.5                    & 20.2                   & 3.0                    & 16.6                      & 24.6                    \\
UserReview-ItemReviewSummary                       & 4.8                         & 0.9                        & 6.3                         & 0.9                        & 6.0                         & 7.8                      & 19.7 & 5.4                    & 20.0                   & 3.0                    & 16.3                      & 24.2                    \\
UserReviewSummary-ItemReview                       & 2.3                         & 0.2                        & 2.7                         & 0.2                        & 2.5                         & 6.0                      & 8.3  & 2.2                    & 8.6                    & 1.1                    & 7.2                       & 23.4                     \\
UserReviewSummary-Item      & 7.0                         & 1.6                        & 8.8                         & 1.6                        & 8.3                         & 10.2                     & 20.3 & 5.4                    & 20.1                   & 3.0                    & 17.0                      & 23.9                 \\\hline   
\end{tabular}
\end{table*}

We are interested in learning the impact of different representation methods for the user and the item.  For user representation, we consider two options:
 1) \textbf{UserReview}:  This composes of the user's reviews of other items written before purchasing the given item. 
2)  \textbf{UserReviewSummary}: This composes of an LLM summarized version of UserReview. To be specific, we induce LLM to summarize the user's purchase reason and post-purchase experiences from the past reviews. 
For product representation, we explore three options:
    1) \textbf{Item}: This is item's metadata includes title and description. 
    2) \textbf{ItemReview}: Based on Item, we additionally incorporate past reviews of this item written by other users before the given user's purchase. 
    3)  \textbf{ItemReviewSummary}: Similar to ItemReview but replace the past reviews with an LLM generated summary.
    
We use Gemini Ultra for this experiment with a zero-shot setup. For past reviews, we select up to 10 of the most recent past reviews, with a maximum limit of 8k tokens. All previous reviews are presented in reverse chronological order. 
As shown in Table~\ref{tab:user_item_representaiton}, representing users with their raw past reviews and items with their metadata is most effective for both generation tasks. Summarizing past user reviews risks losing personal details, while including reviews from other users may introduce irrelevant information.
Further research is needed to extract useful information from such noisy data. In the following experiments, we report the results using UserReview+Item combination.



\subsection{Evaluating Multiple Models (RQ 2)}

\begin{table}[t]
\small
\caption{Performance of all methods on 1k sampled dataset (\%). Gemini is statistically significant better than  other methods via 
\textit{t}-test ($p<0.01$) on all metrics.}
\label{tab:method_compare}
\begin{tabular}{lrrrrrr} \hline
       & \multicolumn{1}{l}{B1}  & \multicolumn{1}{l}{B2}   & \multicolumn{1}{l}{R1}  & \multicolumn{1}{l}{R2}   & \multicolumn{1}{l}{RL}  & BERT \\ \hline
\multicolumn{7}{l}{Purchase reason}                                                                                                               \\
\textbf{Gemini} & 16.5                    & 8.4                      & 21.8                    & 9.2                      & 20.1                    &   25.2   \\
GPT-4 (one-shot)  & 14.3                    & 5.8                      & 18.0                    & 5.4                      & 16.1                    &   19.2   \\
Gemma (one-shot) & 8.2                      & 0.1          & 6.6       & 0.1       & 7.9  & 10.6\\
P5     & 1.2 & 0.06 & 2.7 & 0.06  & 2.6 &   8.5   \\ \hline
\multicolumn{7}{l}{Post-purchase experience}                                                                                                      \\
\textbf{Gemini} & 22.9                    & 8.9                      & 24.1                    & 5.7                      & 20.0                    &    28.5  \\
GPT-4 (one-shot) & 20.1                    & 5.6                      & 21.4                    & 3.1                      & 16.8                    &     23.5 \\
Gemma (one-shot) & 19.6   & 2.5 & 14.5 & 1.6 & 16.6  & 21.8\\
P5     & 0.6 & 0.05 & 4.9 & 0.18 & 4.7 & 7.1     \\ \hline
\end{tabular}
\end{table}

We demonstrate that Gemini Ultra is capable in performing the two generation tasks in a zero-shot setup. 
We additionally benchmark our dataset with GPT-4-Turbo, and a smaller open-sourced Gemma-7B. We also experiment with P5, a pre-LLM state-of-the-art recommendation explanation model for comparison purpose. P5 is a T5-based model trained on five recommendation tasks (e.g., rating prediction, explanation generation) using Amazon reviews. All the comparisons is on a randomly sampled of 1000 reviews from our dataset and Table~\ref{tab:method_compare} show the results.  
We observe that P5 frequently generates short, generic explanations (e.g., \quotes{great product}). This aligns with prior findings~\cite{liu2023} and results in poor performance regardless of whether the explanations are classified as purchase reasons or post-purchase experiences.

\begin{table}[th]
\small
\caption{Performance on 1k sampled dataset with ground truth generated by both Gemini and GPT-4 (\%).  Gemini is statistically significant better than GPT-4 via 
\textit{t}-test ($p<0.01$) on all metrics.}
\label{tab:gpt4_gold_result}
\begin{tabular}{llllllll}
\hline
Ground truth             & Model & B1   & B2   & R1   & R2   & RL   & BERT \\ \hline
\multicolumn{8}{l}{Purchase reason} \\
\multirow{2}{*}{Gemini}                   & Gemini      & 16.5 & 8.4  & 21.8 & 9.2  & 20.1 & 25.2 \\
                  & GPT-4 & 14.3        & 5.8  & 18   & 5.4  & 16.1 & 19.2     \\
\multirow{2}{*}{GPT-4}                    & Gemini      & 15.8 & 6.7  & 21.7 & 6.7  & 19.7 & 21.9 \\
& GPT-4 & 14.6        & 5.2  & 19.6 & 5    & 17.4 & 17.5       \\\hline
\multicolumn{8}{l}{Post-purchase experience} \\
\multirow{2}{*}{Gemini}                    & Gemini      & 22.9 & 8.9  & 24.1 & 5.7  & 20   & 28.5 \\
                  & GPT-4  & 20.1        & 5.6  & 21.4 & 3.1  & 16.8 & 23.5      \\
\multirow{2}{*}{GPT-4 } & Gemini      & 24.5 & 9.3  & 24.5 & 5.1  & 19.9 & 26.8 \\                
& GPT-4 & 23.4        & 6.8  & 23.4 & 3.4  & 17.8 & 22.8 \\

\hline
\end{tabular}
\end{table}

For other LLMs, initially GPT-4 performs significantly worse than Gemini with the same zero-shot prompt. Adding a one-shot example improves GPT-4's performance, but it still  performs worse than Gemini. This raises a  concern: \textbf{is there any ground truth source bias?} In other words, is the performance difference caused by potential biases introduced by using an LLM extractor (Gemini) to create the ground truth?

To answer this, we use GPT-4 to generate the ground truth purchase reasons and post-purchase experiences and re-evaluate performance for both models (Table~\ref{tab:gpt4_gold_result}). Overall Gemini and GPT-4 models are robust and  Gemini consistently has a significant win, regardless of the ground truth source. The specific metric values do vary depending on the ground truth source. This discrepancy stems from the inherent flexibility of language generation and the potential variations in vocabulary and writing style among different LLMs.  Regarding specific tasks, the post-purchase experience  seems more sensitive to the ground truth source: both models achieve higher scores when GPT-4 provides the ground truth. This discrepancy could be attributed to the diverse capabilities exhibited by different LLMs.

Finally, we observe that the smaller Gemma-7B model lags significantly behind the larger models. Future research could explore fine-tuning smaller models to bridge this performance gap.


\subsection{Improving Purchase Reason Prediction (RQ 3) }
\label{subsec:improve_purchase_reason}

We focus on purchase reason prediction and explore how to improve the performance of LLMs on this particular task. We start with some case studies and observe that the model performance is contingent on the presence of relevant information within user's previous reviews. Table~\ref{tab:exp_case_studies} shows two typical examples of good and poor purchase reason generation by Gemini Ultra.  In Example 1, where the user purchased a dog toy, the model accurately deduces the purchase reason is for a large dog since the user has reviewed leash for a 155lb dog before. However, in Example 2, the lack of relevant information in the user history forces the model to rely on product metadata, resulting in a generic purchase reason. This decline in performance is especially noticeable when the actual purchase reason is highly personalized or differs significantly from the product metadata. 

This inspires us to explore improvement from two directions. Firstly, inducing the model to focus more on the past purchases that are relevant to the current product. To this end, we leverage the chain-of-thought (CoT) idea by first asking the model to identify relevant past purchases and then perform the purchase reason generation task. Secondly, incorporating a larger portion of user history might give model richer contextual information (we use up to 10 most recent reviews in the previous experiments). We thus switch Gemini Ultra to Gemini 1.5 Pro~\cite{geminiteam2024}, a recent model with extended context capabilities 
for this exploration. We experiment with up to 32K tokens of past reviews which could hold all the past reviews for 97.33\% of instances. Lastly, as our focus is purchase reason, we remove the post-purchase experience generation to simplify the task.

As from Table~\ref{tab:gemini_pro_result}, we see that the task simplification (\#2) slightly improves the performance over the baseline where purchase reason and post-purchase are jointly generated (\#1). Guiding the model to focus on relevant past purchases (\#3) yields a strong performance boost, improving all metrics by relatively 4-5\%. Conversely, including all past reviews in the history has a detrimental effect (\#4).  This suggests that the model may encounter difficulties identifying relevant past purchases within a lengthy context. This observation is consistent with prior research indicating that LLMs are adept at identifying relevant information at the beginning or end of the input context, but their ability to do so diminishes when the information is located within the middle of a long context~\cite{Liu2024}.










\begin{table}[t]
\centering
\small
\caption{Case studies on purchase reason prediction.}
\label{tab:exp_case_studies} 
\begin{tabular}{p{0.47\textwidth}}
\hline
\textbf{Example 1 (the model generates a good purchase reason)}







\underline{Product title}: HuggleHounds Plush Corduroy Durable Knotties Dog Toy

\underline{Ground-truth purchase reason}: To provide durable chew toys for their dogs, including a large American mastiff.

\underline{Predicted purchase reason}: The user has a large dog and is looking for a durable toy. \\
\underline{Model's rationale}: The user reviewed a heavy-duty leash for a 155lb dog, indicating \textbf{they have a large dog}. The HuggleHounds toy is described as durable and made for dogs of all sizes.\\

\hline \hline
\textbf{Example 2 (the model generates a poor purchase reason)}

\underline{Product title}: QALO Men's Charcoal Step Edge Q2X Silicone Ring Size 08

\underline{Ground-truth purchase reason}: Look for a comfortable ring for work and yard work.

\underline{Predicted purchase reason}: User Anonymous may be looking for a durable and stylish accessory. \\
\underline{Model's rationale}: The user has not left any reviews on similar products, but the product description highlights its durability and style. \\
\hline
\end{tabular}
\end{table}

\begin{table}[ht]
\small
\caption{Gemini 1.5 Pro's results on purchase reason prediction on 10K dataset. */+ denotes statistically better than 1/2 with 
\textit{t}-test ($p<0.01$). }
\label{tab:gemini_pro_result}

\begin{tabular}{llllll}  \hline
  Configuration        & B1   & B2  & R1   & R2  & RL   \\ \hline
1. Baseline             & 19   & 8.5 & 22.8 & 8.1 & 21.0   \\ \hline
2. Purchase reason only & $19.3^*$ & $8.9^*$ & $23.2^*$ & $8.6^*$ & $21.4^*$ \\
3. 2+CoT                & $20.3^{*,+}$ & $9.3^{*,+}$ & $24.1^{*,+}$ & $8.9^{*,+}$ & $22.1^{*,+}$ \\
4. 3+32K token history  & $19.4^{*}$ & $8.8^{*}$ & $23.3^{*}$ & $8.6^{*}$ & $21.3^{*}$ \\ \hline
\end{tabular}
\end{table}

\subsection{Discussion}


We have investigated how various LLMs with prompting perform on the task of predicting purchase reasons and post-purchase experiences. Larger models like Gemini and GPT-4 show similar effectiveness, while smaller models like Gemma lag behind. Future work could explore if fine-tuning these smaller models can bridge this performance gap.

A key challenge is effectively representing user and item information, then prompting the model to identify the most relevant clues for its response. Simply providing a large volume of past reviews makes it difficult to pinpoint the most relevant ones within that extensive context. This highlights the need for better strategies to select pertinent past reviews. Retrieval-augmented generation techniques~\cite{gao2024retrievalaugmented} could be a promising avenue for future exploration.

Our benchmarking evaluation uses lexical and semantic similarity metrics. While informative, they have limitations and do not fully capture the nuances of LLM-generated text. An automated LLM evaluation system could potentially address these limitations. We hope our dataset inspires more research in these directions.

\section{Applications}
\label{sec:applications}


We demonstrate how purchase reason could be beneficial for downstream applications in marketing and recommender systems.

\subsection{Marketing-focused User Behavior Analysis}

Understanding the motivations behind consumer purchases is key to predicting and influencing their behavior. Our dataset and generation method offer a valuable tool for e-commerce sellers to gain insights into why customers choose their products from user reviews. Furthermore, our predictive model can even infer purchase reasons when customers make purchases without leaving reviews.

This analysis can specifically highlight \quotes{gaps} between how a product is described and what customers actually value. Using our curated dataset as a case study, we prompt Gemini Ultra to assess whether the ground truth purchase reasons were adequately captured in the product descriptions. We find that only 33.8\% of purchase reasons were mentioned in the descriptions while the majority 66.2\% are not. Analyzing these uncovered purchase reasons can inspire sellers to refine their product descriptions, bridge the gap between product features and customer needs, and ultimately enhance the customer experience. Future research could explore the integration of purchase reasons into existing product description generation systems ~\cite{Wang2017,Chen2019b}.



\subsection{Building a Better Recommender System}

We experiment how purchase reasons, the initial motives at buying, impact the prediction of post-purchase behaviors (experience and rating) in recommender system. The basic task setup is similar to our purchase reason prediction:  Given a user's past reviews  and item metadata (UserReview-Item representation, detailed in Section~\ref{subsec:user_item_representation}), we prompt Gemini Ultra to generate the two predictions separately. 

\textbf{Post-purchase experience prediction}. We explore variations in both model input (by adding either predicted or actual purchase reasons) and model output (by jointly generating purchase reasons or not). 
Table~\ref{tab:experience_result} shows that including purchase reasons significantly improves model performance across all metrics. Specifically, when purchase reason is jointly predicted, it outperforms the baseline (without purchase reasons) by 5.7\% in BLEU-1, 9.2\% in ROUGE-1, and 10.0\% in ROUGE-L scores.  Providing the ground truth purchase reason (the oracle case) leads to even greater improvements.

\begin{table}[t]
\caption{The impact of purchase reason on post-purchase experience prediction with Gemini Ultra on 10K dataset. URI/Exp/PR denote UserReview+Item representation/Post-purchase experience/Purchase reason, respectively. The differences between each of the methods are statistically significant via \textit{t}-test ($p<0.01)$. }
\label{tab:experience_result}
\small
\begin{tabular}{lllllll} \hline
  Model Input / Output                             & B1   & B2  & R1   & R2 & RL    \\ \hline
1. URI / Exp   & 21.7 & 7.4 & 22.3 &  4.3  & 18.3     \\
2. URI / Exp + PR          & 23.0 & 8.7 & 24.3 & 5.4   & 20.2       \\
3. URI+predicted PR / Exp & 22.0 & 8.0 & 23.5 &  4.9  & 19.4       \\
4. URI+gold PR  / Exp     & 23.3 & 9.3 & 25.1 &  5.8  & 20.7    \\ \hline  
\end{tabular}
\end{table}

\begin{table}[t]
\caption{The impact of purchase reason on rating prediction with Gemini Ultra on 10K dataset. * denotes statistically significant better than baseline (\#1) via a  \textit{t}-test ($p<0.01)$. }
    \label{tab:rating_result_purchase_reason}
    \small
\begin{tabular}{lllc}\hline
    Model input                  & RMSE & MAE   \\\hline
1. UserReview+Item              & 1.59 & 1.02  \\
2. 1+ predicted purchase reason & $1.34^*$ & $0.77^*$  \\
3.  1+ gold purchase reason & $1.31^*$ & $0.72^*$  \\\hline
\end{tabular}
\end{table}


\textbf{Rating prediction}. We experiment with both predicted and actual purchase reasons as additional model input. To evaluate, we use common metrics like Root Mean Square Error (RMSE) and Mean Absolute Error (MAE) to compare predicted and actual ratings. 
As with post-purchase experience prediction, using predicted purchase reason improves rating prediction, reducing RMSE by relatively 15.7\%, MAE by relatively 24.5\%
~(see Table~\ref{tab:rating_result_purchase_reason}). As expected, the upper bound setup of using ground truth purchase reason brings even more improvement. 

\textbf{Explanation}. We highlight that purchase reasons can serve as a novel form of personalized explanation. They reveal the user's initial need when making a purchase and how the products fulfill that need. This perspective complements most existing explanation recommendation methods that focus on justifying items by predicting user sentiment and experience after purchase. A user study or online A/B test would be valuable to evaluate the real-world effectiveness of purchase reasons as recommendation explanations.
\section{Conclusion}
\label{sec:conclusion}

We introduce a novel task of purchase reason prediction, aiming at better capturing what affects a user's decision to purchase a product. We propose an LLM-based approach to generate a high quality, personalized dataset that consists of real-world purchase reasons and post-purchase experiences based on user reviews. Our approach demonstrates strong generalization ability, yielding strong performance across diverse large language models.  As the first of its kind, we showcase the dataset's value by benchmarking it against purchase reason and post-purchase experience prediction tasks. Moreover, we demonstrate the utility of purchase reasons in various downstream applications.  

This new dataset opens up exciting possibilities for building better purchase reason prediction models, particularly by refining user and item representations. Additionally, our study reveals a significant performance gap between smaller and larger models in a few-shot setting. A natural extension is to explore if fine-tuning a smaller model can effectively close this gap. Furthermore, purchase reasons could serve as a novel form of explanation in recommendation systems. It will be valuable to explore the real-world effectiveness of purchase reason explanation via user-study or online A/B test.  We leave these as future work.

\section{Ethical Considerations}




We have created a dataset of purchase reasons and post-purchase experiences based on publicly available Amazon product reviews. User identities in the original Amazon dataset are anonymized, so our dataset does not reveal any private information.

While our methodology for dataset generation can be applied to other review sources, it is important to acknowledge that reviews may not always be genuine or may even be AI-generated. Therefore, we urge caution when utilizing our approach on other review datasets. It is crucial for users, particularly sellers, to be mindful of the potential for inauthentic reviews and to critically assess the quality and reliability of the source material before conducting any analyses.






\bibliographystyle{ACM-Reference-Format}
\bibliography{reference}


\section{Appendix}

This section shows a few prompts used by our LLMs, including
\begin{itemize}
    \item The prompt used for extracting explanations from user reviews (Table~\ref{tab:data_prompt_full}).
    \item The prompt used by LLM auto-rater to evaluate extracted purchase reason (Table~\ref{tab:rater_reason_prompt}) and post-purchase experience (Table~\ref{tab:rater_experience_prompt}).
    \item The prompt to summarize user past reviews (Table~\ref{tab:review_summary_prompt}).
    \item The prompts to generate purchase reason and post-purchase experience in both zero-shot (Table~\ref{tab:benchmark_purchase_reason_prompt_zero_shot}) and one-shot (Table~\ref{tab:benchmark_purchase_reason_prompt_one_shot}) setups.
\end{itemize}

\begin{table*}[t]
    \caption{Prompt used to extract purchase reason and post-purchase experience based on a product and an associated user review.}
    \label{tab:data_prompt_full}
    \small
    \centering
    \begin{tabular}{|p{0.95\linewidth}|}
    \hline
You are a customer engagement specialist at Amazon. Given a product and a customer review, please identify the customer's purchase reason and post-purchase experience:\\
1. explicit\_purchase\_reason:The reason for purchasing the product that is directly mentioned in the review. Describe in detail the thought processes before the purchase. Leave null if not mentioned.\\
2. implicit\_purchase\_reason: The reason for purchasing the product that is not explicitly mentioned in the review but can be inferred from the review and the product description. Leave null if not mentioned.\\
3. purchase\_reason\_explanation: Briefly justify your reasoning behind the identified explicit and implicit purchase reasons.\\
4. post\_purchase\_experience: Summarize in 2-3 lines how the product met (or didn't meet) the customer's expectations, based on the review. \\\\

Please be as specific and relating to the customer's personal context as much as possible.\\ \\

Please answer in valid json format, for example:\\
\{\\
  "explicit\_purchase\_reason": "........",\\
  "implicit\_purchase\_reason": "........",\\
  "purchase\_reason\_explanation": "........",\\
  "post\_purchase\_experience": "........"\\
\}\\
-----------------------------------------------------------------------\\

Here are two examples:

Product:\\
Suncast PB6700 Patio Bench, Light Taupe\\
Suncasts 50 gallon patio bench provides comfortable seating along with convenient storage. It is perfect for storing gardening supplies, patio accessories and more. This patio bench is decorative and functional and will look great in your backyard.\\\\

Customer Review:\\
Good but....\\
I bought this to use as a toy box and seating for my kids. After putting it together which is more tricky than you first think I put toys in it and let my kids play with it then I noticed on the instructions it says not for kids or to be used as a toy box but I'm not sure why. It holds all my kids toys and they love it and as long as I don't let them stand on it or jump on it I think it's perfect. It's not as sturdy as I thought but let's hope it holds up and lasts a while\\\\

Expected Answer:\\
\{\\
"explicit\_purchase\_reason": "To store kids' toys and provide seating for them.",\\
"implicit\_purchase\_reason": "50-gallon storage capacity",\\
"purchase\_reason\_explanation": "The customer needed a solution that could both store toys and serve as a seating area for their children.",\\
"post\_purchase\_experience": "The bench met the customer's expectations for storage and seating. However, it was not as sturdy as they had anticipated."\\
\}\\\\

Product:\\
The World of Beads\\
Muto is a Kodansha International author.\\\\

Customer Review:\\
Five Stars\\
AWESOME AWESOME BOOK..:)\\\\

Expected Answer:\\
\{\\
"explicit\_purchase\_reason": null,\\
"implicit\_purchase\_reason": null,\\
"purchase\_reason\_explanation": "The review does not provide any information about the customer's purchase reasons.",\\
"post\_purchase\_experience": "The customer was very satisfied with the book, describing it as 'AWESOME' and 'Five Stars'."\\
\}\\

-----------------------------------------------------------------------\\
Please analyze purchase reason and post-purchase experience based on this product review and product information. \\\\

Product:\\
\textit{Actual product information including title and description.} \\\\

Customer Review:\\
\textit{Actual review content.} \\\\

Answer:\\
\hline
    \end{tabular}
\end{table*}

\begin{table*}[ht]
    \caption{Prompt used to evaluate purchase reasons based on a product and an associated user review.}
    \label{tab:rater_reason_prompt}
    \small
    \centering
    \begin{tabular}{|p{0.95\linewidth}|}
    \hline
As a customer engagement team leader at Amazon, your task
involves evaluating a summary written by a specialist about why a certain
purchase was made. \\\\
You will analyze the summary based on the provided product information and
customer review, using these criteria: \\

1. completeness: Answer "Yes" or "No". "Yes" if the summary successfully
captures the majority of reasons behind the purchase; otherwise, "No".\\

2. completeness\_reason: Provide a concise explanation for your assessment of
the summary's completeness.\\

3. hallucination: Answer "Hallucination" or "Factual". "Hallucination" if the
summary includes any unrelated details not mentioned or implied by the product
information or the customer review; otherwise, "Factual".\\

4. hallucination\_reason: Provide a concise explanation for your assessment of
the summary's hallucination.\\

5. correctness: Answer "Yes" or "No". "Yes" if the summary exclusively focuses
on pre-purchase information without discussing any post-purchase experiences;
otherwise, "No".\\

6. correctness\_reason: Provide a concise explanation for your assessment of
the summary's correctness.\\
\\

Please respond using a valid json format, for example:\\
\{\\
  "completeness": "Yes",\\
  "completeness\_reason": "...",\\
  "hallucination": "Factual",\\
  "hallucination\_reason": "...",\\
  "correctness": "Yes",\\
  "correctness\_reason": "..."\\
\}\\\\

Now, please evaluate the following summary based on the above criteria:\\\\

Product:\\
\textit{Actual product information.} \\\\

Customer Review:\\
\textit{Actual review content.} \\\\

Specialist's summary of the reasons for purchase: \\
\textit{Actual purchase reasons.} \\\\

Assessment:\\

\hline
    \end{tabular}
\end{table*}

\begin{table*}[ht]
    \caption{Prompt used to evaluate post-purchase experience based on a product and an associated user review.}
    \label{tab:rater_experience_prompt}
    \small
    \centering
    \begin{tabular}{|p{0.95\linewidth}|}
    \hline
As a customer engagement team leader at Amazon, your task
involves evaluating a summary written by a specialist about a user's experience
after purchasing a product.
\\\\
You will analyze the summary based on the provided customer review and product
information, using these criteria:
\\
1. completeness: Answer "Yes" or "No". "Yes" if the summary successfully
captures the majority of the customer's post-purchase experience; otherwise, "No".\\

2. completeness\_reason: Provide a concise explanation for your assessment of
the summary's completeness.\\

3. hallucination: Answer "Hallucination" or "Factual". "Hallucination" if the
summary includes any unrelated details not mentioned or implied by the product
information or the customer review; otherwise, "Factual".\\

4. hallucination\_reason:  Provide a concise explanation for your assessment of
the summary's hallucination.\\\\

Please respond using a valid json format, for example:\\
\{ \\
  "hallucination": "Hallucination", \\
  "hallucination\_reason": "...", \\
  "completeness": "Yes",\\
  "completeness\_reason": "..."\\
\} \\\\

Now, please evaluate the following summary based on the above criteria:\\\\

Product: \\
\textit{Actual product information.} \\\\

Customer Review: \\
\textit{Actual review content.} \\\\

Specialist's summary of post-purchase experience: \\
\textit{Actual post-purchase experience.} \\\\

Assessment: \\

\hline
    \end{tabular}
\end{table*}

\begin{table*}[ht]
    \caption{Prompt used to rewrite user history}
    \label{tab:review_summary_prompt}
    \small
    \centering
    \begin{tabular}{|p{0.95\linewidth}|}
    \hline
This is user Anonymous who left some past reviews on Amazon, please take a look at this user's past review history on other products \\\\

Past reviews from user Anonymous on other products: \\
\textit{User's past reviews} \\\\

Let's identify a few past purchases from this user and predict user's purchase reason for past products and post purchase sentiments \\
1. explicit\_purchase\_reason: why user Anonymous could purchase this past product, as inferred from user's past reviews. \\
2. implicit\_purchase\_reason: why user Anonymous could purchase this past product, not mentioned in user user's past reviews. \\
3. purchase\_reason\_explanation: why do you think this could be the purchase reason \\
4. post\_purchase\_experience: how did this past product meet user Anonymous' expectation, based user user's past reviews, describe it in 2 to 3 lines. \\\\
 
For example: \\

Past item 1: .... \\
\{ \\
  "explicit\_purchase\_reason": "........" \\
  "implicit\_purchase\_reason": "........" \\
  "purchase\_reason\_explanation": "........" \\
  "post\_purchase\_experience": "........" \\
\} \\\\

Past item 2: .... \\
\{ \\
  "explicit\_purchase\_reason": "........" \\
  "implicit\_purchase\_reason": "........" \\
  "purchase\_reason\_explanation": "........" \\
  "post\_purchase\_experience": "........" \\
\} \\\\

Past item 3: .... \\
\{ \\
  "explicit\_purchase\_reason": "........" \\
  "implicit\_purchase\_reason": "........" \\
  "purchase\_reason\_explanation": "........" \\
  "post\_purchase\_experience": "........" \\
\} \\\\

\hline
    \end{tabular}
\end{table*}

\begin{table*}[ht]
    \caption{Zero-shot prompt used for purchase reason and post-purchase experience prediction.}
    \label{tab:benchmark_purchase_reason_prompt_zero_shot}
    \small
    \centering
    \begin{tabular}{|p{0.95\linewidth}|}
    \hline
This is user Anonymous who left some past reviews on Amazon. Please take a look at this user Anonymous' past review history on other products and predict user Anonymous purchase reason for this product and post purchase sentiment \\
1. explicit\_purchase\_reason: why this user could purchase this product, as inferred from this user's past reviews. \\
2. implicit\_purchase\_reason: why this user could purchase this product, not mentioned in this user's past reviews, can be from the product description. \\
3. purchase\_reason\_explanation: why do you think could be the purchase reason. \\
4. post\_purchase\_experience: how could this product meet this user's expectation based this user's past reviews, describe it in 2 to 3 lines. \\ \\

Product:\\
\textit{Actual product information including title and description.} \\\\

Past reviews from this user on other products:\\
\textit{Past review content with past product metadata.} \\\\

Now let's predict this user's purchase reason for this product and post purchase sentiment:
Please answer in json format, for example: \\
\{\\
  "explicit\_purchase\_reason": "........"\\
  "implicit\_purchase\_reason": "........"\\
  "purchase\_reason\_explanation": "........"\\
  "post\_purchase\_experience": "........"\\
\}\\
\\\\
Answer:\\

\hline
    \end{tabular}
\end{table*}

\begin{table*}[ht]
    \caption{One-shot prompt used for purchase reason and post-purchase experience prediction.}
    \label{tab:benchmark_purchase_reason_prompt_one_shot}
    \small
    \centering
    \begin{tabular}{|p{0.95\linewidth}|}
    \hline
This is user Anonymous who left some past reviews on Amazon. Please take a look at this user Anonymous's past review history on other products and predict user Anonymous's purchase reason for this product and post purchase sentiment:\\
1. explicit\_purchase\_reason: why user Anonymous could purchase this product, as inferred from user Anonymous's past reviews. Leave null if you could not find any.\\
2. implicit\_purchase\_reason: why user Anonymous could purchase this product, not mentioned in user Anonymous's past reviews but inferred from the product description. Leave null if you could not find any.\\
3. purchase\_reason\_explanation: why do you think this could be the purchase reason.
4. post\_purchase\_experience: how could the product meet user Anonymous's expectation, based on user Anonymous's past reviews. Please describe in 1 or 2 sentences.\\
\\

Please answer in json format, for example:\\
\{\\
"explicit\_purchase\_reason": "........",\\
"implicit\_purchase\_reason": "........",\\
"purchase\_reason\_explanation": "........",\\
"post\_purchase\_experience": "........"\\
\}\\
\\
--------------------------------------\\
\\
Below is an example from another user.\\
\\

Past product reviewed by this user:\\
\\
Product:\\
Moto E 2nd Gen Case, [Invisible Armor] Xtreme SLIM, CLEAR, SOFT, Lightweight, Shock Absorbing TPU Bumper/ Back Cover for Moto E (2nd Gen, 2015)\\
Why choose Tektide [Invisible Armor]? -Ultra Soft from Inside Out; Ultra Soft Interior prevent your devices getting scratched from the inside. -Skin like Fit; every curve of the case is designed to match your devices to fit it perfectly like skin. -Stays Firmly on Your Device; [Invisible Armor] stays firmly on your device and prevent it from slipping out in any event. -Non Slippery Property; Non Slippery Skin like Material lets you firmly hold your device without your device sliding around. -Access to All Ports; Buttons; All ports, camera and sensors are precisely cut out and pressing buttons requires no extra force. -Premium Quality Guarantee; Hassle Free Replacement Guarantee;  We check carefully for the tiniest defects of our products before shipping to our customers and we offer hassle free replacement by contacting us.\\
User's review:\\
Nice protection, not slippery, fits perfectly.\\
Good protection, clear to show off bamboo back, fit perfectly, inexpensive, overall excellent.\\
User's rating:\\
5\\
\\

Current product information:\\
Moto X Pure Edition Unlocked Smartphone With Real Bamboo, 16GB White Bamboo (U.S. Warranty - XT1575)\\
Stunning 5.7' Quad borderless display on a device that's extremely comfortable to hold and use. Hyper-intelligent 21 MP camera with enhanced focusing technology, zero shutter lag and a colon adjusting flash for the fastest capture and best images. All day battery life (30+ hrs.) that can be turbo powered to give you 10 hours on a 15 min charge. Pure android experience with innovative Motorola enhancement. Supports all major NA carriers. For any troubleshooting, please go to the following link : \url{https://motorola-global-portal.custhelp.com/app/utils/prod_guided_assistant/g_id/4813/p/30,6720,9541/action/auth}\\
\\
Expected answer:\\
\{\\
"explicit\_purchase\_reason": null,\\
"implicit\_purchase\_reason": "To have a large screen and good camera quality with a long battery life.",\\
"purchase\_reason\_explanation": "The product features a large screen, 21 MP camera, and long battery life, which are not mentioned in the user's past reviews but could be reasons for purchasing this product.",\\
"post\_purchase\_experience": "The phone could meet the user's expectations as it offers a large screen, good camera quality, and long battery life. The clear case from the previous review suggests that the user may appreciate the bamboo design of the phone."\\
\}\\
\\
--------------------------------------\\
Below is the information for the current user Anonymous.\\
\\
Past reviews from this user on other products:\\
\textit{Past review content with past product metadata.} \\\\

Current product information:\\
\textit{Actual product information including title and description.} \\\\

Now let's predict user Anonymous's purchase reason and post purchase sentiment for this product.\\\\

Answer:\\

\hline
    \end{tabular}
\end{table*}


\end{document}